\documentclass[12pt]{article}    
\begin{document}

\begin{titlepage}
\begin{flushright}
                Preprint IFT UWr 932/2000\\
                May 2000 \\ 
                hep-th/0005097
\end{flushright}
\bigskip

\begin{center}
{\Large \bf
Light-cone formulation \\
and \\
spin spectra
of
massive strings\\}
\end{center}
\bigskip

\begin{center}
        {\bf Marcin Daszkiewicz\footnote{
                E-mail: marcin@ift.uni.wroc.pl}\\and\\
         {Zbigniew Jask\'{o}lski}
           \footnote{
                E-mail: jaskolsk@ift.uni.wroc.pl; fax: (48 71) 3201 409}} 
                          \bigskip\\
        Institute of Theoretical Physics, University of Wroc{\l}aw, 
        \\
        pl.Maxa Borna 9, 50-204 
        Wroc{\l}aw, Poland\\
\end{center}    

\bigskip \bigskip

\begin{abstract}                                   
It is shown that  all  bosonic
and fermionic massive string models  admit  
consistent light-cone formulations.  This result is used
to derive the spin  generating functions of these models 
in four dimensions.

\end{abstract}
\bigskip\bigskip



\thispagestyle{empty}
\end{titlepage}

\section{Introduction}  

The covariant quantization of the free Nambu-Goto string \cite{brower72},
and the Ramond-Neveu-Schwarz fermionic string \cite{schwarz72,brower73}  
leads in non-critical 
dimensions to  consistent quantum models with longitudinal 
excitations. 
It was first pointed out by Polyakov \cite{polyakov81a,polyakov81b}
that the  dynamics of these extra degrees of freedom  should be 
described by the (super)-Liouville theory. 
This suggests the  modifications of the standard  string world-sheet 
actions by adding the (super)-Liouville
sectors. In the context of free string such
modifications were first analyzed by Marnelius \cite{marnelius}. 
More recently the covariant quantization
of these models was reconsidered under the assumption 
that the bulk and the boundary cosmological constants vanish 
\cite{haja,hajaos}. The analysis of the physical state conditions 
in terms of DDF operators exhibits three types of excitations
corresponding to the transverse,
the Liouville, and the longitudinal degrees of freedom. 
The first two are described by the standard Fock spaces 
while the latter one is given by 
a unitary highest weight representation of the (super)-Virasoro algebra
\cite{haja,hajaos}.
The classification of the ghost-free models is then
related to the classification of the unitary (super)-Virasoro Verma 
modules \cite{gokeol}, 
and yields one continuous, and one discrete series of 
models.
In all cases the first excited state is massive
which justifies the name {\it massive string} introduced
in our previous papers.

One member of the discrete series is especially interesting.
It corresponds to the (super)-algebra of longitudinal excitations
with vanishing central charge. In this case  
the  bosonic massive string
is equivalent to the non-critical Nambu-Goto string \cite{haja}, 
and the fermionic one  to the non-critical RNS string
\cite{hajaos}. Both models admit the 
quantum light-cone formulation which can be used to 
analyse their spin spectra \cite{dahaja98,dahaja99}.

In the present letter we address the problem of
the structure and the physical content 
of other quantum massive string models. 
Our main result is the light-cone formulation 
of all  models.
We present the derivation for
the discrete series, but the same technique 
can be used for the continuous series as well. 
The light-cone formulation allows for a complete description of 
the spin content of the theory. This is 
done in terms of the spin generating functions  
for the bosonic and fermionic massive strings in four
dimensions. 

Our results show that 
all the massive string models 
have essentially the same structure. In contrast to the critical
strings where the number of the tachion-free models is strongly limited, 
the GSO projection  yields  a large class of 
tachion-free massive strings 
 with surprisingly rich mass and spin spectra. 
None of these models contains massless states with spin
greater than 1. This makes them   good candidates for matter fields
in an effective description of low energy QCD. 
The central problem of such
application   is  a consistent interaction. 
There are at least two possibilities: either the contact
joining-splitting interaction, or the exchange of a vector particle
to which the massive string naturally couples by its world-sheet action.
The light-cone formulation provides a setting in which one can 
pose and in principle answer the question of the Lorentz covariance 
of both kinds of interactions. 
This was  our main motivation for the present work.

The paper is divided in two parts. The first one is devoted to the 
bosonic models and contains  more detailed derivations.
The second part concerns the fermionic strings.

\section{Bosonic string}
We define the light-cone massive bosonic string  as a
representation of the algebra
$$
\begin{array}{l}
\begin{array}{rllrllrl}
{[a_0^i,q_0^j]}\!\!\!
&=\; - i \delta^{ij}\;
&,
&\;{[a_0^+,q_0^-]}\!\!\!
&=\;i\;
&,
&\;[c_0,q_0^L]\!\!\!
&=\;-i\;\;\;,\\[6pt]
{[a^i_m,a^j_n]}\!\!\!
&=\; m \delta^{ij} \delta_{m,-n}\;
&,
&\;[c_m,c_n]\!\!\!
&=\; m \delta_{m,-n}\;
&,
\end{array}\\[22pt]
{[L^{\rm \scriptscriptstyle L}_m,L^{\rm \scriptscriptstyle L}_n]}
\;=\; (m-n) L^{\rm \scriptscriptstyle L}_{m+n} + {c^{\rm
\scriptscriptstyle L}\over 12} (m^3 - m) \delta_{m,-n} \;\;\;.
\end{array}
$$
All other commutators vanish and the standard conjugation
properties  are assumed.

The algebra of non-zero modes is by construction
isomorphic to the (diagonalized)
algebra of the DDF operators of the covariant approach
\cite{haja}.   In particular, the algebra of $L^{\rm
\scriptscriptstyle L}_m$ corresponds to the Virasoro algebra of
the shifted Brower longitudinal operators with the central charge
$c^{\rm \scriptscriptstyle L}$. The value of 
$c^{\rm \scriptscriptstyle L}$ is restricted by the no-ghost theorem 
\cite{haja} to the range $1\leq c^{\rm \scriptscriptstyle L}< 25-d$,
or to the discrete series
$$
 c^{\rm \scriptscriptstyle L}=
c_m\equiv 1 - {6\over m(m+1)} \;\;\;,\;\;\;m=2,3,... \;\;\;\;. 
$$
Let us denote by ${\cal V}_m(p,q)$ the unitary Verma module of
the Virasoro algebra with the central charge $c_m$ and with  the
highest weight \cite{gokeol} 
$$
h_m(p,q)  =
{((m+1)p -m q)^2 - 1 \over 4m(m+1)}
\; \;\;,\;\;\; 1\leq p\leq m-1;\;\; 1\leq q\leq p\;\;.
$$
The  space of states in the light-cone formulation is given by
$$
{\cal H}_m(p,q)=
\int {dp_+ \over | p^+|}
 d^{d-2}\overline{p}\;
 {\cal F}(p^+,\overline p)\otimes
 {\cal V}_m(p,q)\;\;\;,
$$
where ${\cal F}(p^+,\overline{p})$ denotes  the Fock space
generated by the transverse and the Liouville excitations 
out of the unique vacuum state $\Omega$ satisfying $$
\sqrt{\alpha}a^i_0\,\Omega = p^i \,\Omega\;\;\;,\;\;\;
\sqrt{\alpha}a^+_0\,\Omega = p^+
\,\Omega\;\;\;,\;\;\; c_0\,\Omega = 0\;\;\;. $$
In order to construct a unitary realization
of the Poincare algebra on ${\cal H}_{m}(p,q)$ we introduce
\begin{eqnarray*}
L_{n}^{\rm \scriptscriptstyle A} &=&
 {\textstyle\frac{1}{2}}\sum_{k= -\infty}^{+\infty}
 \sum_{i=1}^{D-2}
 :a_{-k}^ia_{n+k}^i:
 \;\;\; \nonumber \\
 L_{n}^{\rm \scriptscriptstyle C} &=&
 {\textstyle\frac{1}{2}}\sum_{k= -\infty}^{+\infty}
 :c_{-k}c_{n+k}: \;+\;2i\sqrt{\beta} n c_n \;+\; 
 2\beta\delta_{n,0}
 \;\;\;. \nonumber          
\end{eqnarray*}
The operators $L^{\rm \scriptscriptstyle T}_n\equiv 
L_{n}^{\rm \scriptscriptstyle
A} +  L_{n}^{\rm \scriptscriptstyle C}
+L_{n}^{\rm \scriptscriptstyle L}$
form the Virasoro algebra with the central charge
$
c_m^{\rm \scriptscriptstyle T} = d + 48\beta   -
{\textstyle{6\over m(m+1)}}
$.
Following the standard light-cone construction
we define the Poincare generators
\begin{eqnarray*}
P^+&=&\sqrt{\alpha}a^+_0\;\;\;,\;\;\; P^i=\sqrt{\alpha}a^i_0
\;\;\;,\;\;\;{P}^- =
{\frac{\sqrt{\alpha}}{a^+_0}}
(L^{\rm \scriptscriptstyle T}_0 - \alpha_0) \;\;\;,\\
{M}^{ij}_{\rm lc\;\;} &=& {P}^i{x}^j-{P}^j{x}^i +
i\sum_{n\geq 1}
{\frac{1}{n}} (a_{-n}^ia_{n}^j - a_{-n}^ja_{n}^i)
\;\;\;,
\nonumber \\
{M}^{i+}_{\rm lc\;\;} &=& {P}^+ {x}^i \;\;\;,\nonumber\\
{M}^{+-}_{\rm lc\;\;} &=& {\textstyle\frac{1}{2}}({P}^+{x}^-+x^-P^+) \;\;\;,
\nonumber \\
{M}^{i-}_{\rm lc\;\;} &=& {\textstyle\frac{1}{2}}( {x}^i P^-
 + P^- {x}^i ) - {P}^i{x}^- - 
 i{\frac{ \sqrt{\alpha}}{P^+}} \sum_{n\geq 1} {\frac{1}{n}}
\left( a_{-n}^i L_n^{\rm \scriptscriptstyle T}
- L_{-n}^{\rm \scriptscriptstyle T} a_n^i \right), \nonumber
\end{eqnarray*}
It is known \cite{dahaja98,schwarz} that the algebra of these
generators closes to the Lie algebra of Poincare group if and only
if the central charge $c_m^{\rm \scriptscriptstyle T} $ of the
Virasoro algebra  $\{ L_n^{\rm \scriptscriptstyle T} \}$ and
$\alpha_0$ entering the definition of $P^-$ take the critical
values $ c_m^{\rm \scriptscriptstyle T} = 24,\;
\alpha_0 = 1. $ This conclusion is independent of any
detailed structure of $L_n^{\rm \scriptscriptstyle T}$ as far as
the commutation relations 
$$ [L_m^{\rm \scriptscriptstyle T},
a^i_n ] = -n a^i_{m+n} \;\;\;,\;\;\; [L_m^{\rm \scriptscriptstyle
T}, q^i_0 ] = -i a^i_m \;\;\;, 
$$ 
are satisfied. Solving the condition $c_m^{\rm
\scriptscriptstyle T} = 24$ for the parameter $\beta$ one recovers
the discrete series of values predicted by the no-ghost theorem of
the covariant approach \cite{haja}: $$ \beta = \beta_m\; =
 \;{24-d\over 48} + {1\over 8m(m+1)}
 \;\;\;.
$$
Calculating the mass square operator one gets
\begin{equation}
\label{mass}
M^2 = 2\alpha\left( R_m(p,q) + {((m+1)p - m q )^2 \over 4m(m+1)}
 - {d\over 24}
\right)
\end{equation}
where  $R_m(p,q)$
is the level operator in ${\cal H}_m(p,q)$. 

For any $m\geq 2$ the space of states ${\cal H}_m(p,q)$ is by
construction isomorphic to the space ${\cal H}^{\rm cov}_m(p,q)$
of physical states of the
corresponding covariant massive string model. It follows from
(\ref{mass}) that the mass spectra of these models coincide.  
In the case of $m=2$  also the Poincare algebra representations are 
isomorphic \cite{dahaja98}. In order to extend this result for all $m>2$ 
we shall  construct an isomorphism
${\cal I}:{\cal H}^{\rm cov}_m(p,q) \rightarrow {\cal H}_m(p,q)$.
To this end it is convenient to work with the parameterization of 
${\cal H}^{\rm cov}_m(p,q)$ in terms of the "original"  (not diagonal) 
basis of the transverse $A^i_{n}$, the Liouville $C_{n}$, and 
the not "shifted" longitudinal $\widetilde{A}^{\scriptscriptstyle 
L}_{n}$ DDF operators. (For details concerning the DDF 
construction and the notation used in this section we refer to 
\cite{dahaja98}.) Since all states in ${\cal H}^{\rm cov}_m(p,q)$ are 
on mass-shell, $x^+$ 
can be regard as an evolution parameter
in any representation in which the operator 
$\hat{x}^+$ is diagonal. Each 
physical state $|\psi\rangle \in {\cal H}^{\rm cov}_m(p,q)$ is then  
uniquely determined by its $x^+=0$ initial condition 
$\overline{|\psi\rangle}$, and each operator $A$ on ${\cal H}^{\rm 
cov}_m(p,q)$ defines an operator $\overline{A} \:\overline{|\psi\rangle} 
\equiv \overline{A|\psi\rangle}$ acting on the space 
$\overline{{\cal H}}_m(p,q)$ of all $x^+=0$ initial conditions. 

We define the isomorphism ${\cal I}:{\cal H}^{\rm cov}_m(p,q) 
\rightarrow {\cal H}_m(p,q)$
as a composition
${\cal I} = {\cal J}\circ\overline{{\cal I}}$, where
$\overline{{\cal I}}:{\cal H}^{\rm cov}_m(p,q) \ni
|\psi\rangle \rightarrow \overline{|\psi\rangle}\in 
\overline{{\cal H}}_m(p,q)$,
and ${\cal J}:
\overline{\cal H}_m(p,q)
\rightarrow {\cal H}_m(p,q)$ is given by
\begin{equation}
\label{sub}
{\cal J} \overline{A}^i_n {\cal J}^{-1} = a^i_n \;\;\;;\;\;\;
{\cal J} \overline{\widetilde A}^{\scriptscriptstyle L}_n
{\cal J}^{-1} = L^{\scriptscriptstyle T}_n
\;\;\;;\;\;\;
{\cal J} \overline{C}_n {\cal J}^{-1} = c_n \;\;\;.
\end{equation}
The   Poincare algebra representations
are equivalent if for all covariant  Poincare generators $G_{\rm cov}$
and their light-cone counterparts $G_{\rm lc}$ the following
relations hold
\begin{eqnarray}
\label{con1}
{\cal J} \overline{[G_{\rm cov}, A^i_n] }{\cal J}^{-1}
 &= &  [G_{\rm lc}, a^i_n] \;\;\;,\\
\label{con2}
{\cal J}
\overline{ [G_{\rm cov}, \widetilde{A}^{\scriptscriptstyle L}_n]}
{\cal J}^{-1}
&= &
[G_{\rm lc}, L^{\scriptscriptstyle T}_n] \;\;\;,  \\
\label{con3}
{\cal J}\overline{[G_{\rm cov}, C_n]}{\cal J}^{-1}
&= &[G_{\rm lc}, c_n] \;\;\;.
\end{eqnarray}
The Poincare generators    
as well as the transverse $A^i_n$, and the longitudinal
$\widetilde{A}^{\scriptscriptstyle L}_n$ operators are identical in all
covaraint models. Also the light-cone commutators appearing on the
right hand sides of the equations (\ref{con1}), and (\ref{con2}) 
coincide. Indeed, for all $m>2$   the algebra of
$L^{\scriptscriptstyle T}_n$, and $ a^i_n$ on ${\cal H}_m$ is isomorphic
to the corresponding one on ${\cal H}_2$.
Thus, since the equations (\ref{con1},\ref{con2}) are satisfied for $m=2$
\cite{dahaja98} so they are for all $m>2$.

The Liouville  DDF operator $C_n$ depends on $m$ only via the
constant $\beta_m$ and similar arguments can be used to
prove the third equation. It can be also directly
calculated for all $m$ by means of the leading term method 
\cite{goreth}.
The only non-vanishing commutator on the l.h.s. of (\ref{con3}) reads
\begin{eqnarray*}
\overline{\left [{M}^{i-}_{\rm cov},C_n \right ]} & =&
-\alpha\frac {n x^i}{p^+} \overline{C}_n
+ \sqrt\alpha \frac {in}{p^+} \left(\sum_{m>0} \frac {1}{m}
(\overline{A}_{-m}^i \overline{C}_{n+m} - \overline{A}_m^i\overline{C}_{n-m})
\right)                                               \\
&  &-{\sqrt\alpha}\frac {2}{p^+} n \sqrt {\beta_m} \overline{A}_n^i \;\;\;,
\end{eqnarray*}
and after substitution (\ref{sub}) it reproduces the light-cone commutator
$[M^{i-}_{\rm lc},c_n]$. This completes our proof  that 
the light-cone models introduced in this section
provide  the light-cone formulations of the covariant models
found in \cite{haja}. As the simplest application of this result
we shall  investigate the spin content of these models in four dimensions.

For every $m\geq 2$ there are ${1\over 2} m(m-1)$ ground states.
For $m>2$ there are always tachionic and massive ground states.
Massless ground states are very rare.
The first one appears at $m=24,p=q=20$ the next at $m=242,p=q=198$.

For all $m\geq 2$ and allowed $p,q$ the first 
excited states are massive and the space
$
{\cal F}(p^+,\overline{p})\otimes {\cal V}_m(p,q)
$                                    
carries a nonlinear representation of the little group SO(3)
and a linear representation of its "transverse" subgroup SO(2).
The generating function for SO(2) characters
takes the form
\begin{equation}
\label{gener}
\chi^m_{p,q}(t,\varphi) =t^{- h_m(p,q)} \chi^m_{p,q}(t)
\chi(t)  \prod\limits_{k\in I\!\!N}
{1\over 1- 2t^k \cos(\varphi) + t^{2k} }
\end{equation}
where $ I\!\!N$ is the set of positive integers.
The  partition function
$$
\chi(t) = \prod\limits_{k\in I\!\!N} (1- t^k)^{-1}\;\;\;
$$
counts states in the Liouville sector.  
The multiplicity  of states in the longitudinal sector
is given by the minimal model partition function \cite{ro}
$$
\chi^m_{p,q}(t) = \chi(t)
\sum\limits_{k\in Z\!\!\!Z}  
\left( t^{\alpha^m_{p,q}(k)} -  t^{\alpha^m_{p,-q}(k)} \right)\;\;\;, 
$$                                                                 
where
$$
{\alpha^m_{p,q}(k)} = {[2m(m+1)k -q m + p(m+1) ]^2 -1 \over 
4m(m+1)}\;\;\;.  
$$
Taking into account the decomposition 
$
\chi^j(\varphi) =  1+ \sum_{k=1}^{j} 2\cos(k\varphi)
$ of
the irreducible SO(3) character 
$\chi^j(\varphi)$ of spin $j\in I\!\!N$
into irreducible characters of the SO(2) subgroup,
one can obtain the SO(3) generating function
expanding  (\ref{gener}) directly in terms
of $\chi^j(\varphi)$. Using formulae (5.3) and (5.4) of Ref.\cite{cuth}
one gets
$$
\chi^m_{p,q}(t,\varphi) =    \sum\limits_{j\in I\!\!N_0}
\chi^m_{p,q}(j,t) \chi^j(\varphi)\;\;\;,
$$
where $I\!\!N_0$ is the set of non-negative integers and 
the spin $j$ generating function $\chi^m_{p,q}(j,t)$ is given by
$$
\chi^m_{p,q}(j,t) =   t^{- h_m(p,q)}
\chi^m_{p,q}(t) \chi(t)^3
\sum\limits_{k\in I\!\!N}
t^{{k(k-1)\over 2}}(-1)^{k-1}(1-t^k)^2t^{kj}
\;\;\;.
$$
The expression above provides a complete description of the SO(3)
representations. Note that the full space reflection has a
non-linear realization on the space
${\cal F}(p^+,\overline{p})\otimes {\cal V}_m(p,q)$. 
This makes the
analysis of the O(3) content of the model much more complicated.

The construction of the light-cone formulation of the closed 
string models proceeds along the standard lines by tensoring two 
open string spaces of states with an appropriate 
identification of the zero modes, and with the additional  
constraint $(L_0^{{\scriptscriptstyle T}\,{\rm (left)}} - 
L_0^{{\scriptscriptstyle T}\,{\rm (right)}}) |\psi\rangle =0$ imposed. 
For 
$m=2,3,4$ the only states satisfying this constraint belong to the 
"diagonal" sectors 
$
{\cal H}_m^{\rm (left)}(p,q) \otimes {\cal H}_m^{\rm (right)}(p,q)
$.
The spin content can be easily deduced from the diagonal part
(i.e. all terms of the form $x^ay^a$) of
$\chi^m_{p,q}(x,\varphi)\chi^m_{p,q}(y,\varphi) $.

For $m\geq 5$ one gets also non-diagonal solutions corresponding
to pairs  $(p,q)\neq (p',q')$ such that the difference
$
h_m^{\rm (left)}(p,q)- h_m^{\rm (right)}(p',q')
$
is an integer number. For $m=5$  such pairs are given by
$h_5(4,1)- h_5(1,1)= 3$ and $h_5(3,1)- h_5(2,1)= 1$.
The spin content of the corresponding models can be easily derived
from modified products of generating functions. For instance
in the case  of
${\cal H}_5^{\rm (left)}(4,1)\otimes {\cal H}_5^{\rm (right)}(1,1)$
 the spin content is given  by the diagonal
part of the product
$\chi^5_{4,1}(x,\varphi)y^3\chi^5_{1,1}(y,\varphi) $.
For $m>5$ the number of non-diagonal models slowly increases
with $m$.

\section{Fermionic string}

We define  the discrete series of fermionic non-critical light-cone string
models as representations of the algebras of the transverse
$$
\begin{array}{rllrllrl}
{[a_0^i,q_0^j]}\!\!\!
&=\; - i \delta^{ij}\;
&,
&\;{[a_0^+,q_0^-]}\!\!\!
&=\;i\;
&,\\[6pt]
{[a^i_m,a^j_n]}\!\!\!
&=\; m \delta^{ij} \delta_{m,-n}\;
&, &\;\{b^i_r,b^j_s \}\!\!\!
&=\; \delta^{ij} \delta_{r,-s}\;
&,
\end{array}
$$
the Liouville
$$
\begin{array}{rllrllrlll}
&\;[c_0,q_0^L]\!\!\!
&=\;-i\;
&, &\; {[c_m,c_n]}\!\!\!
&=\; m \delta_{m,-n}\;
&, &\;\{d_r,d_s \}\!\!\!
&=\; \delta_{r,-s}\
&,
\end{array}
$$
and the longitudinal
$$
\begin{array}{l}
{[L^{\rm \scriptscriptstyle L}_m,L^{\rm \scriptscriptstyle L}_n]}
\;=\; (m-n) L^{\rm \scriptscriptstyle L}_{m+n} + {\hat{c}_m^{\rm
\scriptscriptstyle L}\over 8} (m^3 - m) \delta_{m,-n} \;\;\;,\\[10pt]
\left[ L^{\rm \scriptscriptstyle L}_m, G^{\rm \scriptscriptstyle L}_r \right]
\; = \;({\textstyle\frac{1}{2}}m - r)G^{\rm \scriptscriptstyle L}_{m+r}
   \;\;\;,\\[10pt]
\left\{ G^{\rm \scriptscriptstyle L}_r, G^{\rm \scriptscriptstyle L}_s \right\}
 \;= \; 2L^{\rm \scriptscriptstyle L}_{r+s}
                + {\textstyle{
                \hat{c}_m^{\rm \scriptscriptstyle L}
                \over 2}}(r^2-
{\textstyle{1\over 4}})\delta_{r,-s}\;\;\;
\end{array}
$$
excitations, supplemented by the standard conjugation properties. In 
the formulae above 
$m,n\in Z\!\!\!Z;r,s \in Z\!\!\!Z + {\textstyle{\epsilon\over 2}}$,
and $\epsilon = 1$ corresponds to the Neveu-Schwarz sector,
while $\epsilon = 0$ to the Ramond sector. The construction 
depends on the discrete parameter  $m=2,3,...$ via the central charge
$$
\hat{c}_m^{\rm \scriptscriptstyle L} = 1 - {8\over m(m+2)}
\;\;\;
$$
of the superconformal algebra  of the longitudinal
excitations.
Let us denote by  ${\cal V}_m^\epsilon(p,q)$
the unitary Verma module  with the highest weight  \cite{gokeol}
$$
h^\epsilon_m(p,q) = {((m+2)p - m q)^2 - 4 \over 8m(m+2) }
+ {1-\epsilon \over 16}\;\;\;
$$
where for the Neveu-Schwarz sector ($\epsilon =1$) 
$1\leq q \leq p \leq m-1$,  and $p-q$ is even. For the Ramond sector 
($\epsilon =0$) $p-q$ is odd, and 
  $1\leq q \leq p-1$ for $1\leq p \leq [{1\over 2} (m-1)]$,
and   $1\leq q \leq p+1$ for $ [{1\over 2} (m+1)] \leq p \leq 
m-1$. We assume that
${\cal V}_m^0(p,q)$ is a representation of the extended
Rammond algebra which includes a fermion chirality operator $(-1)^F$.
The highest weight vectors are thus two-fold degenerate, unless
$h^0_m(p,q) - {\hat{c}^{\rm \scriptscriptstyle L}\over 16} = 0$.
Let ${\cal F}_\epsilon(p^+,\overline{p})$
be the Fock space generated by the algebra of non-zero
transverse and Liouville excitations
out of the vacuum state   $\Omega_\epsilon$ satisfying $$
\sqrt{\alpha}a^i_0\,\Omega_\epsilon = p^i \,\Omega_\epsilon\;\;\;,\;\;\;
\sqrt{\alpha}a^+_0\,\Omega_\epsilon = p^+
\,\Omega_\epsilon\;\;\;,\;\;\; c_0\,\Omega_\epsilon = 0\;\;\;. $$
In the Neveu-Schwarz sector ($\epsilon = 1$)
the space of states of the light-cone model
is defined by
$$
{\cal H}^1_m(r,s) = \!\int {dp_+ \over | p^+|} d^{d-2}\overline{p}\;
{\cal F}^1(p^+,\overline p)\otimes
 {\cal V}^1_m(r,s)\;\;\;.
$$
In the Rammond sector ($\epsilon = 0$) the fermionic zero modes
$b^i_0, d_0$ along with the 0-level fermion chirality operator
$\Gamma^F$
form the real Euclidean Clifford algebra ${\cal C}(d,0)$.
Let $D(d)$ be the irreducible representation of the complexified Clifford
algebra ${\cal C}^C(d) = {\cal C}(d,0)\otimes C\!\!\!\!I$.
We define the space of states in the Ramond sector ($\epsilon = 0$)
by
$$
{\cal H}^0_m(p,q) = \int {dp_+ \over | p^+|} d^{d-2}\overline{p}\;
{\cal F}^0(p^+,\overline p)\otimes D(d)\otimes
 {\cal V}^0_m(p,q)\;\;\;,
$$
where the tensor product of the operator algebras representations 
is determined by the requirement that the subspace of
non-excited states provides an irreducible
representation $D(d+1)$ of the
complexified Clifford algebra
${\cal C}^C(d+1)$ generated by $b^i_0, d_0, \Gamma^F$ and 
$G_0$. For  even $d$,
$D(d+1) = D(d)$, and  the construction of 
${\cal H}^0_m(p,q)$ does not depend on 
whether the highest weight vector of $ {\cal V}^0_m(p,q)$
is degenerate or not.

The construction of  a unitary representation of
the Poincare group is a straightforward generalization 
of the construction given in Ref. \cite{dahaja99} for the
fermionic critical ($m=2$) massive string.  We introduce the operators
\begin{eqnarray*}
  L_{m}^{\rm \scriptscriptstyle A}   & = & 
  {\textstyle{1\over 2}}\!\! \sum\limits_{n \in Z\!\!\!Z} 
  \sum_{i=1}^{D-2}     
  :{a}^i_{-n} {a}^i_{n+m}: \;+\;
{\textstyle{1\over 2}} \!\!\!
 \sum\limits_{r \in Z\!\!\!Z + {\epsilon\over 2} }
 \sum_{i=1}^{D-2}     
   r \,:{b}^i_{-r} {b}^i_{r+m}: \;+\;
   (1-\epsilon) {\textstyle{d-2\over 16}}\delta_{m,0}\nonumber
   \;\;\;,\\
L_{n}^{\rm \scriptscriptstyle C} & = &
{\textstyle{1\over 2}}\!\! \sum\limits_{n \in Z\!\!\!Z}
: c_{-n}c_{n+m}:\; +\;
{\textstyle{1\over 2}} \!\!\!\sum\limits_{r \in Z\!\!\!Z +
{\epsilon \over 2} }
 r\, :d_{-r} d_{r+m}:\;+
 2i\sqrt{\beta}m c_m +
 [ {\textstyle{1-\epsilon\over 16}}+2\beta]\delta_{m0}\;, \\
G_r^{\rm \scriptscriptstyle A}   & = &
 \sum\limits_{n \in Z\!\!\!Z}
 \sum_{i=1}^{D-2}     
{a}^i_{-n} {b}^i_{n+r}\;\;\;, \\
G_r^{\rm \scriptscriptstyle C}   & = &  
 \sum\limits_{n \in Z\!\!\!Z}
c_{-n}d_{n+r}\; + \;4i\sqrt{\beta} r d_r \;\;\;.\nonumber
\end{eqnarray*}  
The operators      
$L_m^{\rm \scriptscriptstyle T}=
L_m^{\rm \scriptscriptstyle A} +
L_m^{\rm \scriptscriptstyle C} +
L_m^{\rm \scriptscriptstyle L}$, and        
$G_r^{\rm \scriptscriptstyle T}=
G_r^{\rm \scriptscriptstyle A} +
G_r^{\rm \scriptscriptstyle C} +
G_r^{\rm \scriptscriptstyle L}$                                  
form an $N=1$ superconformal algebra with the central charge
$
\hat{c}^{\rm \scriptscriptstyle  T} = 
\hat{c}_m^{\rm \scriptscriptstyle L}+ d-1+32\beta
$.
The generators of the Poincare group are defined
by
\begin{eqnarray*}  
P^+&=&\sqrt{\alpha}a^+_0\;\;\;,\;\;\; P^i=\sqrt{\alpha}a^i_0
\;\;\;,\;\;\;  
{P}^- =
{\frac{\alpha}{P^+}} (L_0 - \alpha_0) \;\;\;  \\
M^{ij}& = &  x^i P^j - x^j P^i -
i\sum\limits_{n>0}
{{1\over n}} ( a^i_{-n} a^j_{n} - a^j_{-n} a^i_{n}
)\;\;\;\nonumber\\
&&\hspace{10pt}
+\; (1- \epsilon) i b^i_0 b^j_0
- i\sum\limits_{r>0}
( b^i_{-r} b^j_{r} - b^j_{-r} b^i_{r}) \;\;\;,  \nonumber\\
M^{i+}&=& x^iP^+\;\;\;,\\
M^{+-}& =& {\textstyle{1\over 2}} (P^+x^- + x^-P^+)\;\;\;,\\
M^{i-}& = &  {\textstyle{1\over 2}} (P^-x^i + x^iP^-) - x^- P^i -
{i\over a^+_0}\sum\limits_{n>0}
{{1\over n}} ( a^i_{-n} L^{\rm \scriptscriptstyle T}_{n} - 
L^{\rm \scriptscriptstyle T}_{-n} a^i_{n}
)\;\;\;\nonumber\\
&&\hspace{10pt}
+\; (1- \epsilon) {i\over a^+_0} b^i_0 G^{\rm \scriptscriptstyle T}_0- 
{i\over a^+_0}\sum\limits_{r>0}
( b^i_{-r} G^{\rm \scriptscriptstyle T}_{r} - 
G^{\rm \scriptscriptstyle T}_{-r} b^i_{r}
)  \;\;\;.
\end{eqnarray*}   
This algebra closes if and only if   
$\hat{c}^{\rm \scriptscriptstyle  T} = 8$
and $\alpha_0= {1\over 2}$ \cite{dahaja99,schwarz}. Solving these 
conditions one gets the discrete series of values of the parameter $\beta$
$$
\beta = \beta_m = {8-d\over 32} + {1\over 4m(m+2)}
\;\;\;,\;\;\;m=2,3,...\;\;\;,
$$ 
and the mass square operator    
$$
M^2_\epsilon = 2\alpha\left( R^\epsilon_m(p,q) 
+ {((m+2)p - m q )^2 \over 8m(m+2)}
 - {d\over 16}\epsilon
\right)
$$
where  $R^\epsilon_m(p,q)$  
is the level operator in ${\cal H}^\epsilon_m(p,q)$.
This exactly coincides with the results of 
the covariant approach \cite{hajaos}.
Following the method of the previous section and using the 
results of Ref.\cite{dahaja99} one can show that in each sector 
and for every admissible $m,p,q$
the representations of the Poincare algebra in the light-cone model
and in the covariant model are isomorphic.
This completes the derivation of 
the light-cone formulation of the discrete series of the
fermionic massive models obtained in \cite{hajaos}.
We shall briefly consider their  spin content  in four dimensions. 

In the Neveu-Schwarz  sector all ground states  are scalars.
For $m=2$ there is one tachionic ground state.
For $m>2$ there are always tachionic and massive ground states.
The massless ground states are very rare, the first one appears at
$m=16, p=q=12$, the next at $m=98, p=q=70$. 

In the Ramond sector all ground states are massive,
except the only massless  cases of even $m$ and  
$p={m\over 2}, q= {m\over 2} +1$.
In the massive case the space of ground states carries two massive
spin ${1\over 2}$ representations while in the massless case
there are two pairs of left and right Weyl spinors.

The generating function for SO(2) characters (normalized to 
integer and half-integer powers) 
takes the form               
\begin{eqnarray*}
\chi_{p,q}^{\epsilon, m}(t,\varphi)  &=&
t^{-h^{\epsilon}_m(p,q)}
\chi_{p,q}^{\epsilon,m }(t)  \chi^{\epsilon}(t)        
\left( 4\cdot 2 \cos {\varphi\over 2}\right)^{1-\epsilon}      
 \\
&\times & 
\prod\limits_{k\in I\!\!N}
{1\over 1- 2t^k \cos(\varphi) + t^{2k} }  
\prod\limits_{r\in I\!\!N-{\epsilon\over 2}}
( 1 + 2t^r \cos(\varphi) + t^{2r} )  \;\;\;,
\end{eqnarray*}   
where the partition function
$$
\chi^\epsilon(t) = 
 \prod\limits_{k\in I\!\!N} {\;1+ t^{k -{\epsilon\over 
2}}
\over 1- t^k}\;\;\;
$$
counts states in the super-Liouville sector.  
The multiplicity  of states in the longitudinal sector
is given by the character of the unitary representation 
of the super-Virasoro algebra (not extended in the Ramond sector)
 \cite{gokeol}
$$
\chi^{\epsilon,m}_{p,q}(t) =  \chi^\epsilon(t)   
\sum\limits_{k\in Z\!\!\!Z}  
\left( t^{\gamma^m_{p,q}(k)} -  t^{\gamma^m_{-p,q}(k)} \right)\;\;\;, 
$$                                                                 
where
$$
{\gamma^m_{p,q}(k)}= {[2m(m+2)k -p (m+2) + q m ]^2 -4 \over 
8m(m+2)}\;\;\;.
$$
Using the techniques developed in Ref.\cite{cuth} one can expand
$\chi_{p,q}^{\epsilon, m}(t,\varphi)$
in terms of the irreducible SO(3) characters $\chi^j(\varphi)$  
$$
\chi_{p,q}^{\epsilon, m}(t,\varphi)     
= \sum_{j \in I\!\!N_0+{1-\epsilon\over 2}}
\chi_{p,q}^{\epsilon,m}(j,t) \chi^j(\varphi )
\;\;\;.
$$        
In the Ramond sector $j \in I\!\!N-{1\over 2}$, and the spin 
generating functions read
\begin{eqnarray*} 
\chi_{p,q}^{0,m}(j,t)&=& 
2t^{-{1\over 8}}
 t^{-h^{0}_m(p,q)}
\chi_{p,q}^{0,m }(t)  \chi^{0}(t)  
p(t)^3      \\
&\times&\!\!\!\!\!\!\! 
\sum\limits_{r\in I\!\!N-{1\over 2}}
\sum\limits_{k\in I\!\!N}
(-1)^{k-1} t^{{\scriptscriptstyle {1\over 2}} (r^2 +k(k-1))}
(1-t^k)(1-t^{r +{\scriptscriptstyle {1\over 2}}} )
(t^{k|j-r|} - t^{k(j+r+1)})      
\end{eqnarray*}  
where in the case of $h^0_m(p,q)=0$ the multiplicity of 
$0$-level, spin ${1\over 2}$ representations should be corrected by 2.                                         
The GSO projection in this sector removes half of the 
representations 
at each excited level \cite{hajaos}.

In the Neveu-Schwarz sector $j \in I\!\!N$, and
\begin{eqnarray*} 
\chi_{p,q}^{1,m}(j,t)&=& 
t^{-h^{1}_m(p,q)}
\chi_{p,q}^{1,m }(t)  \chi^{1}(t)  
p(t)^3      \\
&\times&\!\!\!\!\!\!\! 
\sum\limits_{r\in I\!\!N}
\sum\limits_{k\in I\!\!N}
(-1)^{k-1} t^{{\scriptscriptstyle {1\over 2}} (r^2 +k(k-1))}
(1-t^k)(1-t^{r +{\scriptscriptstyle {1\over 2}}} )
(t^{k|j-r|} - t^{k(j+r+1)})      
\end{eqnarray*}
The GSO projection removes all integer levels. The spin content
of the resulting model can be read of from coefficients 
in front of the  half-integer powers in $\chi_{p,q}^{1,m}(j,t)$.

The fermionic closed string Hilbert space can be constructed 
as the subspace of the tensor product 
${\cal H}^{({\rm left})\epsilon}_m(p,q)\otimes
{\cal H}^{({\rm right})\widetilde{\epsilon}}_{\widetilde{m}}
(\widetilde{p},\widetilde{q})$          
of two open string Hilbert 
spaces determined by the conditions  
$$
a^i_0 = \widetilde{a}^i_0 = {P^i_c\over 2\sqrt{\alpha}}\;\;\;,
\;\;\;
a^+_0 = \widetilde{a}^+_0 = {P^+_c\over 2\sqrt{\alpha}}\;\;\;,
\;\;\;c_0 = \widetilde{c}_0 =0 \;\;\;,
$$
and anihillated by the twist operator  
$L_0^{{\scriptscriptstyle T}\,{\rm (left)}} - 
L_0^{{\scriptscriptstyle T}\,{\rm (right)}}$.                               
Although more general  heterotic-like constructions are possible we
shall restrict ourselves to the models with $m=\widetilde{m}$. We
also assume 
the GSO projection in the left, and in the right sector, separately.                                                                   

Let  $m\geq 2$ and $\epsilon=\widetilde{\epsilon}$. In this case 
the space-time spectrum contains only integer spins. Possible 
models  are restricted by the twist condition
$$
|h^{\epsilon}_m(p,q)-h^{\epsilon}_m(\widetilde{p},\widetilde{q})|
\in I\!\!N_0
\;\;\;
$$ 
which admits obvious diagonal solutions $p=\widetilde{p},q =\widetilde{q}$.
One can check by numerical calculations that  
for sufficiently large $m$ there are 
also some non-diagonal solutions in both sectors.

The half-integer spin spectrum is possible only for the "mixed" 
sectors $\epsilon +\widetilde{\epsilon}=1$. In this case the twist 
condition takes the form 
$$
|h^{1}_m(p,q) + {1\over 2} -
h^{0}_m(\widetilde{p},\widetilde{q})|
\in I\!\!N_0
\;\;\;
$$         
where the factor ${1\over 2}$ comes from the GSO projection. 
One can easily show that there are no solutions for odd $m$. 
Numerical calculations 
suggest that there are always some solutions for even $m$, but we
have not succeeded  in finding a simple proof. 

As in the case of bosonic string the closed string character generating 
function can be obtained as the diagonal part of the product
of the left, and the right open string generating functions.

\end{document}